\documentclass[twocolumn,english,nofootinbib, prl]{revtex4}
\usepackage[T1]{fontenc}
\usepackage[latin9]{inputenc}
\usepackage{amsmath}
\usepackage{amssymb}
\usepackage{graphicx}

\makeatletter

\providecommand{\tabularnewline}{\\}

\@ifundefined{textcolor}{}
{%
 \definecolor{BLACK}{gray}{0}
 \definecolor{WHITE}{gray}{1}
 \definecolor{RED}{rgb}{1,0,0}
 \definecolor{GREEN}{rgb}{0,1,0}
 \definecolor{BLUE}{rgb}{0,0,1}
 \definecolor{CYAN}{cmyk}{1,0,0,0}
 \definecolor{MAGENTA}{cmyk}{0,1,0,0}
 \definecolor{YELLOW}{cmyk}{0,0,1,0}
}

\makeatother

\usepackage{babel}
\begin{document}

\title{How long-range interactions tune the damping in compact stars}

\author{Kai Schwenzer}

\address{Department of Physics, Washington University, St. Louis, Missouri,
63130, USA}
\begin{abstract}
Long-range interactions lead to non-Fermi liquid effects in dense
matter. We show that, in contrast to other material properties, their
effect on the bulk viscosity of quark matter is significant since
they shift its resonant maximum and can thereby change the viscosity
by many orders of magnitude. This is of importance for the damping
of oscillations of compact stars, like in particular unstable r-modes,
and the quest to detect signatures of deconfined matter in astrophysical
observations. We find that, in contrast to neutron stars with standard
damping mechanisms, compact stars that contain ungapped quark matter
are consistent with the observed data on low mass x-ray binaries.
\end{abstract}
\maketitle
Quantum chromodynamics (QCD) features strong, long-range gauge interactions
that lead to color confinement. At high density these interactions
are partly screened by the medium and can lead to deconfinement and
various hypothetical forms of quark matter \cite{Alford:2007xm}.
In the color-flavor locked (CFL) phase at asymptotically large density
all quark modes are gapped, but many of the possible phases at moderate
densities that might be reached in compact stars feature ungapped
quark excitations. Yet, in-medium effects cannot entirely screen the
gluonic interaction, but their magnetic part is only Landau-damped
\cite{Kapusta}. These long-range correlations strongly modify the
low-energy fermionic excitations of deconfined dense matter and lead
to non-Fermi liquid (NFL) behavior \cite{Schafer:2004zf,Schafer:2005mc}
of various material properties \cite{Heiselberg:1993cr,Gerhold:2004tb,Schafer:2004jp,Pal:2011ve}.
Analogous NFL effects due to long-range interactions are studied extensively
in condensed matter systems \cite{stewart2001non} and with dual holographic
descriptions \cite{Iqbal:2011ae}. Whereas in such systems NFL effects
are generally only found in controlled experiments at low temperatures,
we discuss here a case where they have important observable consequences
outside of a physical laboratory and in an environment hotter than
the surface of the sun.

Compact stars are the only known objects that are dense enough that
they could contain deconfined quark matter, and their mechanical oscillation
modes are the only known way to directly probe the dense matter in
their interior. The connection is established via the damping of these
modes which is locally described by the viscosities of the particular
microscopic form of matter. Whereas the shear viscosity is generally
very similar in hadronic and quark matter, since both arise from unscreened
plasmon-enhanced scattering of relativistic particles, namely leptons
\cite{Shternin:2008es} and ungapped quarks \cite{Heiselberg:1993cr},
respectively, the bulk viscosity can be decisively different for various
forms of matter \cite{Sawyer:1989dp,Madsen:1992sx,Alford:2006gy,Alford:2007rw,Alford:2010gw,Alford:2011df}.
In dense matter bulk viscosity arises generally from slow flavor-changing
weak interactions. The bulk viscosity is a resonant mechanism that
features a sharp maximum when the time scale of the relevant weak
process matches the one of a driving density oscillation. Although
the bulk viscosity is induced by weak processes, their rate depends
on strong interactions since they can open phase space for weak transitions.
Correspondingly low energy strong interactions control the resonance,
analogous to the dial of a radio tuner, and even moderate changes
can alter the viscosity at a given temperature and frequency by many
orders of magnitude. 

In this work we study the influence of NFL interactions on the bulk
viscosity of dense quark matter. The bulk viscosity stems from a lag
between a driving density oscillation and the chemical equilibration
of the system mediated by slow weak interactions. Here we study the
damping of small amplitude mechanical oscillations in the subthermal
case where the displacement of the chemical potentials $\mu_{\Delta}$
from their equilibrium values satisfies $\mu_{\Delta}\ll T$. The
suprathermal regime \cite{Madsen:1992sx,Alford:2010gw} and the influence
of strong interactions on the saturation of unstable modes \cite{Alford:2011pi}
will be discussed elsewhere. Whereas in dense hadronic matter chemical
equilibration is caused by semi-leptonic Urca processes, in dense
quark matter strangeness-changing non-leptonic processes $d\!+\! u\leftrightarrow u\!+\! s$
dominate \cite{Madsen:1992sx}. In this case the displacement $\mu_{\Delta}=\mu_{d}-\mu_{s}$
is finite when density oscillations drive the system out of chemical
equilibrium and causes a net non-leptonic rate $\Gamma_{nl}^{\left(\leftrightarrow\right)}\!\equiv\!\Gamma_{d\to s}\!-\!\Gamma_{s\to d}$
that re-equilbrates the system. To compute this rate we follow the
analysis performed in \cite{Heiselberg:1992bd} and extend it to the
interacting case. The computation requires the dispersion relations
of low energy quarks which is modified by self-energy corrections
from quantum fluctuations involving unscreened magnetic gluons. The
form of the dispersion relation including such NFL effects is close
to the Fermi surface given by \cite{Brown:2000eh,Schafer:2004zf,Schafer:2004jp}

\begin{equation}
p_{i}\approx p_{Fi}+v_{Fi}^{-1}\!\left(\!\left(\!1\!+\!\sigma\log\!\left(\!\frac{\Lambda}{E_{i}\!-\!\mu_{i}}\!\right)\!\right)\left(E_{i}\!-\!\mu_{i}\right)\!-\!\delta\mu_{i}\!\right)\label{eq:NFL-dispersion}
\end{equation}
where in the weak coupling, hard dense loop \cite{Braaten:1989mz}
approximation the parameters are given by \cite{Gerhold:2004tb,Schafer:2004jp}

\begin{align}
 & v_{F}\approx1-\frac{2\alpha_{s}}{3\pi}=\quad,\quad\delta\mu\approx\frac{4\alpha_{s}}{3\pi}\mu\label{eq:parameters}\\
 & \sigma\approx\frac{4\alpha_{s}}{9\pi}\quad,\quad m^{2}\approx\frac{2\alpha_{s}}{3\pi}\mu^{2}\quad,\quad\Lambda\approx0.28m\nonumber 
\end{align}
The strong correlations significantly increase the density of states
near the Fermi surface where the group velocity of the corresponding
excitations gradually deviates from its Fermi liquid value $v_{F}$
and eventually vanishes in the low energy limit. This effect arises
from a kinematic low energy enhancement and is even present in the
weak coupling regime. Yet, the parametric form of the NFL dispersion
relation has been shown to hold beyond perturbation theory \cite{Schafer:2004zf}
and via a manifest power counting analysis it has been pointed out
that this result holds within an effective theory in the low temperature
limit even at strong coupling \cite{Schafer:2005mc}. Aside from the
strong generic increase of the density of states close to the Fermi
sea due long-range interactions, the other strong interaction corrections
to the Fermi liquid parameters are not logarithmically enhanced and
due to our ignorance on the values of these parameters in the strong
coupling regime relevant for compact stars we neglect these moderate
corrections here. As discussed in \cite{Schafer:2004jp} there are
likewise no logarithmic corrections to the weak interaction vertices. 

In the subthermal limit $\mu_{\Delta}\ll T\ll\mu$ and neglecting
the subleading quark masses $m_{i}\ll\mu$ we find for the non-leptonic
rate

\begin{equation}
\Gamma_{nl}^{\left(\leftrightarrow\right)}\approx-\frac{64G_{F}^{2}\sin^{2}\!\theta_{c}\cos^{2}\!\theta_{c}}{5\pi^{3}}\mu_{q}^{5}T^{2}\left(\!1\!+\!\sigma\log\!\left(\!\frac{\Lambda}{T}\!\right)\!\right)^{4}\mu_{\Delta}\label{eq:non-leptonic-rate}
\end{equation}
This result differs from the non-interacting expression given in \cite{Madsen:1992sx,Heiselberg:1992bd}
by the strong interaction corrections in terms of NFL factors $\lambda\equiv1\!+\!\sigma\log\!\left(\Lambda/T\right)$
for each of the involved four quarks. These factors become large in
the ultradegenerate regime $T\ll\Lambda=O\!\left(\mu\right)$ realized
in compact stars. In eq. (\ref{eq:non-leptonic-rate}) additional
terms suppressed in powers of $\sigma/\lambda$ which could not be
evaluated in closed form have been neglected, but a numerical integration
shows that their combined contribution is at the relevant temperatures
less than $5\%$.

For comparison we will also study the viscosity due to Urca processes
including NFL interactions. The direct Urca rate is obtained as a
generalization of the Fermi liquid expression \cite{Sa'd:2007ud,Alford:2011df}.
In case of the dominant $d$-quark Urca processes $d\to u+e^{-}+\bar{\nu}_{e}$,
$u+e^{-}\to d+\nu_{e}$ where $\mu_{\Delta}\!=\!\mu_{d}\!-\!\mu_{u}\!-\!\mu_{e}$
we find for the NFL dispersion relation eq. (\ref{eq:NFL-dispersion})

\begin{equation}
\Gamma_{dU}^{\left(\leftrightarrow\right)}\approx\frac{17G_{F}^{2}\cos^{2}\!\theta_{c}\alpha_{s}}{15\pi^{2}}\mu_{q}^{2}\mu_{e}T^{4}\left(\!1\!+\!\sigma\log\!\left(\!\frac{\Lambda}{T}\!\right)\!\right)^{2}\!\mu_{\Delta}\label{eq:Urca-rate}
\end{equation}
As observed previously in the study of the neutrino emissivity \cite{Schafer:2004jp},
this expression exhibits a similar but weaker NFL enhancement since
only two quarks are present in the weak process. The same holds for
the Cabbibo suppressed $s$-Urca channel. However, these Urca rates
are strongly suppressed in $T/\mu$ compared to the non-leptonic rate
eq. (\ref{eq:non-leptonic-rate}).

The bulk viscosity is a measure for the local dissipation caused by
a (harmonic) oscillation of the conserved baryon density $n(\vec{r},t)=\bar{n}(\vec{r})+\Delta\! n(\vec{r})\sin\!\left(\omega t\right)$
and can be defined in terms of the energy dissipation by
\[
\zeta=\frac{2}{\omega^{2}}\left\langle \frac{d\epsilon}{dt}\right\rangle _{{\rm \! diss}}\frac{\bar{n}^{2}}{\left(\Delta\! n\right)^{2}}\,.
\]
Its detailed derivation is for instance discussed in \cite{Alford:2010gw}.
Generally the non-leptonic and the Urca channels mix and require a
coupled-channel analysis \cite{Sa'd:2007ud,Shovkovy:2010xk}, but
due to the smallness of the Urca rate this effect is negligible and
we evaluate them here individually. In the subthermal limit, denoted
by the superscript ``$<$'', the bulk viscosity has the general
resonant form \cite{Alford:2010gw}

\[
\zeta^{<}=\frac{C^{2}\gamma}{\omega^{2}+\left(B\gamma\right)^{2}}=\zeta_{max}^{<}\frac{2\omega B\gamma}{\omega^{2}+(B\gamma)^{2}}
\]
where the resonant maximum $\zeta_{max}^{<}=C^{2}/(2\omega B)$ is
independent of the weak rate $\gamma\equiv\Gamma^{\left(\leftrightarrow\right)}/\mu_{\Delta}$,
but the latter determines at which temperature this maximum is reached.
Here the strong susceptibilities $B$ and $C$ determine the response
of the medium and are given by

\begin{equation}
C\equiv\bar{n}\left.\frac{\partial\mu_{\Delta}}{\partial n}\right|_{x}\quad,\quad B\equiv\frac{1}{\bar{n}}\left.\frac{\partial\mu_{\Delta}}{\partial x}\right|_{n}\label{eq:susceptibilities}
\end{equation}
where $x\equiv n_{\Delta}/n$ is the fraction of the particle density
corresponding to $\mu_{\Delta}$ that is driven out of chemical equilibrium.
The susceptibilities are given for the different processes in table
\ref{tab:strong-parameters} and in particular they do not involve
logarithmic corrections. This can be seen from the pressure at finite
temperature and density which has been computed in weak coupling beyond
leading order in \cite{Ipp:2006ij}. This result shows that all logarithmic
temperature corrections are further suppressed in $T/\mu$. This is
generally expected since in the pressure and the susceptibilities
the entire Fermi sea contributes and the contribution from the thin
shell of width $O\!\left(T\right)$ around the Fermi surface is subleading.

\begin{table}
\begin{tabular}{|c|c|c|c|}
\hline 
 & $\mu_{\Delta}$ & $B$ & $C$\tabularnewline
\hline 
quark non-lept. & $\mu_{d}\!-\!\mu_{s}$ & $\frac{2\pi^{2}}{3\mu_{q}^{2}}$ & $-\frac{m_{s}^{2}}{3\mu_{q}}$\tabularnewline
\hline 
quark d-Urca & $\mu_{d}\!-\!\mu_{u}\!-\!\mu_{e}$ & $\frac{2\pi^{2}}{3\mu_{q}^{2}}$ & $-\frac{m_{d}^{2}-m_{u}^{2}}{3\mu_{q}}$\tabularnewline
\hline 
hadronic Urca & $\mu_{n}\!-\!\mu_{p}\!-\!\mu_{e}$ & $\frac{8S}{n}\!+\negthinspace\frac{\pi^{2}}{\left(4\left(1\!-\!2x\right)S\right)^{2}}$ & $4\!\left(1\!-\!2x\right)\!\left(n\frac{\partial S}{\partial n}\!-\!\frac{S}{3}\right)$\tabularnewline
\hline 
\end{tabular}

\caption{\label{tab:strong-parameters}Strong interaction parameters, defined
in eq. (\ref{eq:susceptibilities}), describing the response of the
particular form of dense matter. For comparison we also show the values
for interacting hadronic matter in terms of the proton fraction $x$
and the symmetry energy $S$. The corresponding weak rate for hadronic
Urca processes is given in \cite{Sawyer:1989dp}. }
\end{table}

Let us first consider the expected size of the NFL corrections. From
our knowledge of the running of the strong coupling \cite{Bogolubsky:2009dc}
at scales $O\!\left(\mu\right)$, quark matter at neutron star densities
is, analogous to the strongly correlated fluid observed at RHIC, expected
to be strongly coupled $\alpha_{s}=O\!\left(1\right)$. Using the
perturbative expressions eq. (\ref{eq:parameters}) as a guide, the
relevant range of the parameters eq. (\ref{eq:parameters}) in compact
stars should be roughly $0.05\lesssim\sigma\lesssim0.3$ and $100\lesssim\Lambda/T\lesssim10^{6}$
so that the enhancement factor $\lambda^{4}$ can take values in the
range $2\lesssim\lambda^{4}\lesssim1000$. Whereas the logarithmic
form eq. (\ref{eq:NFL-dispersion}) is generic, the perturbative relations
between the parameters $\sigma$ and $\Lambda$ and the coupling do
not need to hold, but due to our ignorance of the actual dependence
in the strong-coupling regime we extrapolate them below.

\begin{figure}
\includegraphics{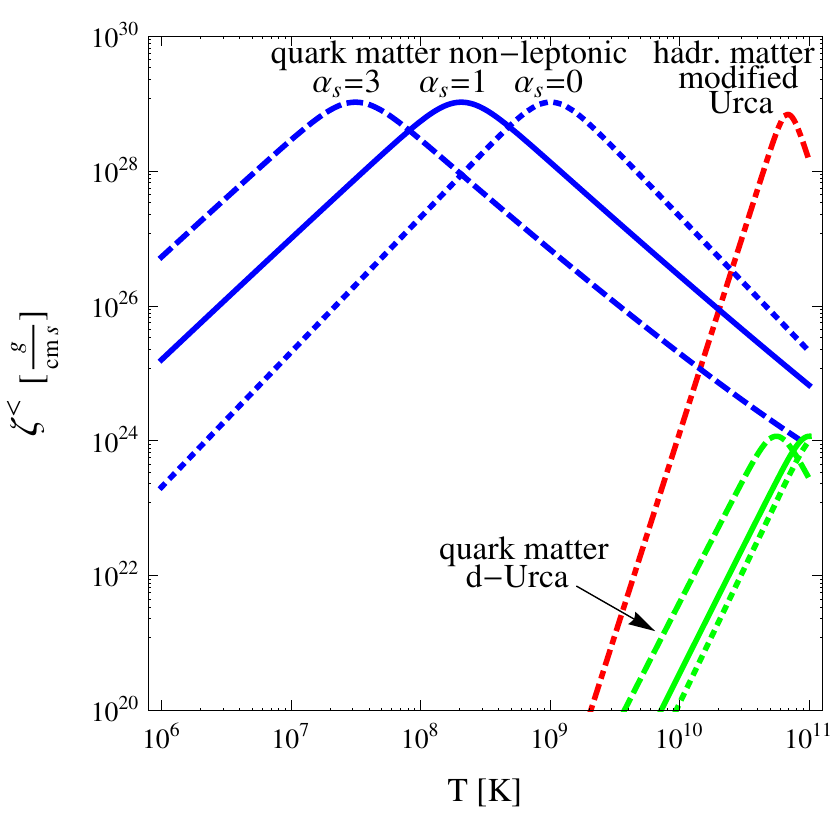}

\caption{\label{fig:bulk-viscosity}The influence of NFL effects due to long-ranged
magnetic gauge interactions on the bulk viscosity of dense quark matter.
All curves are given for a baryon density $n\!=\!3\, n_{0}$ and a
frequency $\nu\!=\!1\,{\rm kHz}$. Shown is the dominant contribution
due to non-leptonic processes for three values of the strong coupling
constant ($\alpha_{s}\!=\!0$ dotted, $\alpha_{s}\!=\!1$ solid and
$\alpha_{s}\!=\!3$ dashed), as well as the subleading contribution
from $d$-quark Urca reactions. For comparison the bulk viscosity
due to modified Urca interactions in dense APR neutron star matter,
studied in \cite{Alford:2010gw}, is shown as well.}
\end{figure}
Our results for the bulk viscosity are shown in fig. \ref{fig:bulk-viscosity}
for different values of the strong coupling, using an intermediate
value for the strange quark mass $m_{s}=150\,{\rm MeV}$. As can be
seen the increase of the non-leptonic rate due to long-range gluonic
interactions shifts the resonant maximum of the bulk viscosity to
significantly lower temperatures and thereby strongly increases the
damping of cold quark matter. As observed before in \cite{Sa'd:2007ud,Shovkovy:2010xk}
the Urca contribution is only relevant at small frequencies and temperatures
$T\!>\!10^{9}\,{\rm K}$. Due to the moderate change of the Urca rate
this conclusion is unaltered by NFL effects. Compared to the corresponding
bulk viscosity in hadronic matter the damping of interacting quark
matter is more than ten orders of magnitude larger at temperatures
$T<10^{8}\,{\rm K}$ present in cores of old compact stars.

\begin{figure}
\includegraphics{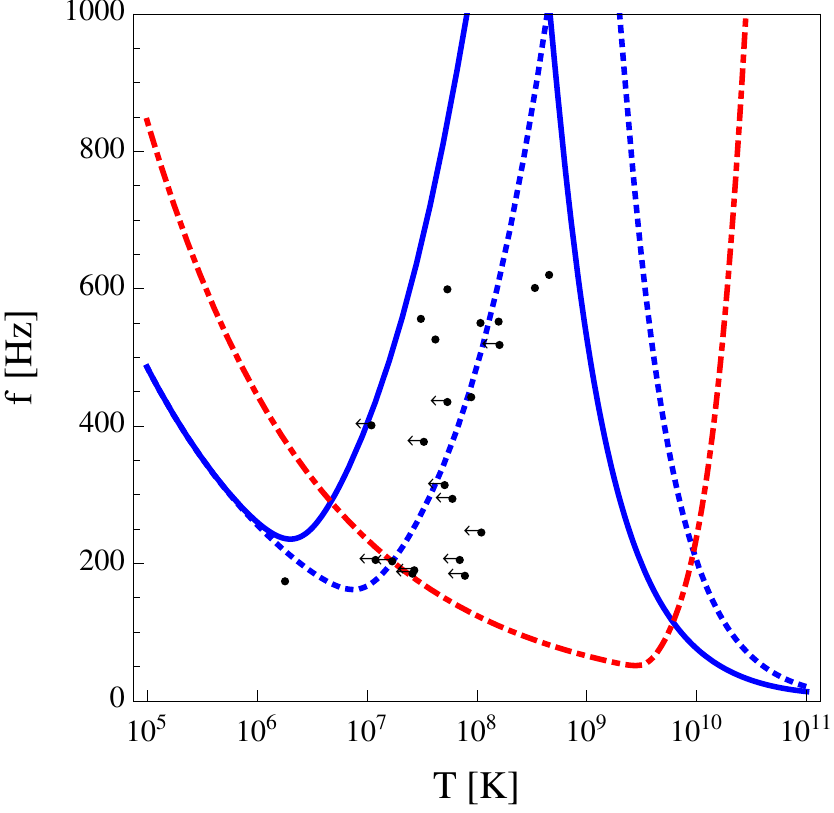}

\caption{\label{fig:instability-region}Impact of the NFL corrections to the
bulk viscosity on the r-mode instability region for a $1.4\, M_{\odot}$
strange quark star. The latter is shown for two values of the strong
coupling constant ($\alpha_{s}\!=\!0$ dotted and $\alpha_{s}\!=\!1$
solid). For comparison the instability region of a $1.4\, M_{\odot}$
neutron star is shown by the dot-dashed curve. For details on the
star models see \cite{Alford:2010fd}. The theoretical curves are
compared to LMXB pulsar data compiled in \cite{Haskell:2012}, where
either temperature estimates are known (points) or only upper bounds
were possible (points with arrows).}
\end{figure}

As an example for the importance of the considered strong interaction
corrections to the bulk viscosity of quark matter we study the instability
of r-modes \cite{Andersson:1997xt,Andersson:2000mf} of strange quark
stars \cite{Witten:1984rs}. R-modes are global oscillation modes
of stars rotating at a finite angular velocity $\Omega$, that are
unstable to the emission of gravitational waves within a characteristic
instability region in the $T$-$\Omega$-plane. At the boundary of
the instability region the gravitational instability is precisely
balanced by viscous damping whereas for frequencies above the r-mode
grows exponentially, for more details see \cite{Alford:2010fd}. If
r-modes are unstable they strongly radiate gravitational waves and
would quickly spin down a star \cite{Owen:1998xg}. Therefore, if
the r-mode growth is stopped at values predicted by known saturation
mechanisms \cite{Bondarescu:2007jw,Alford:2011pi}, observed low mass
x-ray binaries (LMXBs), that are slowly spun up over long time periods
by mass transfer from a companion, should not lie in the r-mode instability
region. Yet, as can be seen in fig. \ref{fig:instability-region}
most observed LMXBs lie clearly in the instability region of a $1.4\, M_{\odot}$
neutron star with standard damping mechanisms, i.e. shear viscosity
due to leptons \cite{Shternin:2008es} and bulk viscosity due to modified
Urca reactions \cite{Sawyer:1989dp}. The situation is even worse
for heavier stars and this conclusion cannot be changed by the involved
uncertainties due to a remarkable insensitivity to the underlying
parameters \cite{Alford:2010fd}. The enhanced dissipation of phases
with ungapped quarks strongly increases the damping at low temperatures
and leads to a stability window where r-modes are stable up to large
frequencies. Nevertheless, neglecting interactions in the bulk viscosity
many observed LMXBs are still within the instability region of a $1.4\, M_{\odot}$
strange star, as shown by the dotted curve. However, taking into account
long-range gauge interactions in the strong coupling case $\alpha_{s}\!=\!1$
all observed pulsars are consistent with lying outside of the instability
region. Therefore, the enhanced damping of ungapped quark matter can
explain the large spin rates of LMXBs.

The results found here for strange stars immediately generalize to
hybrid stars, where a quark matter core is enclosed in a hadronic
outer part. Since the component with the stronger viscosity dominates
the damping the instability region of a hybrid star is roughly given
in each case by the upper segments of the neutron star and strange
star curves \cite{Alford:2010fd}. Further our results apply to certain
color superconducting phases that feature ungapped quark species and
unscreened gluonic modes that can induce a NFL enhancement, see \cite{Schafer:2004jp}.
However, in hyperonic stars \cite{Lindblom:2001hd} where similar
non-leptonic processes exist, there is no NFL enhancement since long
range color interactions are absent.

It is striking that there are no data points for pulsars spinning
with frequencies of several hundred Hertz that at the same time have
core temperatures $\lesssim10^{7}\,{\rm K}$. A stability window,
as found here for ungapped quark matter, provides a robust explanation
for this observation since the strong dissipative heating \cite{Alford:2012yn}
due to r-modes in the instability region prevents stars from entering
it. Clearly, our results involve considerable uncertainties on the
detailed strong coupling values of the various low energy parameters
entering the bulk viscosity. Nevertheless, within the uncertainties
interacting quark matter can provide a consistent explanation for
the the astrophysical data, whereas hadronic matter with standard
damping mechanisms can not \cite{Alford:2010fd}. Potential enhanced
damping mechanisms in hadronic stars related to the complicated neutron
star crust have been proposed, like boundary layer rubbing or mutual
friction in hadron superfluids, see e.g. \cite{Haskell:2012}. It
requires detailed further study to decide if these mechanisms can
likewise explain the astrophysical data, or if the data might eventually
provide a robust signature for deconfined quark matter. In this endeavor
the connection of the dynamic star evolution to pulsar timing data
\cite{Alford:2012yn} should provide important information to distinguish
between these different scenarios. 
\begin{acknowledgments}
I am grateful to Mark Alford and Thomas Schaefer for helpful discussions.
This work was funded in part by the U.S. Department of Energy under
contracts \#DE-FG02-91ER40628 and \#DE-FG02-05ER41375.

\bibliographystyle{h-physrev}
\bibliography{cs}
\end{acknowledgments}

\end{document}